# CURRENT COSMOLOGICAL SIMULATIONS, ADVICE ON THEIR INTERPRETATION, AND AN IMAX FILM PREVIEW

F J SUMMERS
Princeton University Observatory, Peyton Hall, Princeton, NJ 08544;
summers@astro.princeton.edu

ABSTRACT   This general talk is geared toward non-specialists and addresses three questions: (1) What types of cosmological simulations are done? (2) What are the relevant problems to address and what results can one expect? (3) How can one tell if the results are reasonable?

## 1. THE SCALE OF THE PROBLEM

The universe is huge. No, it's more like gigantically humongous. Or maybe it is better described as hyper–super–mega–incomparably beyond any dimension we could ever imagine. And really, it's bigger than that, too. I sure wouldn't want to have to paint it.

In cosmology, researchers have the audacity to attempt numerical simulations of the universe. Of course, one can only address a limited range of scales, usually referred to as the dynamic range of a simulation. The choice to focus upon certain objects or processes fixes the central scales in the numerical model. The push for larger models is to provide a wider dynamic range, perhaps extending orders of magnitude above and below the central scale. The goal is to fit as much as possible of the relevant scales into a finite computational domain.

The extent of the desired domain poses quite a challenge. From superclusters and voids and walls to clusters and groups of galaxies and down to giant through dwarf galaxies, cosmology attempts to explain structure formation on scales from kpc to hundreds of Mpc. The time scales involved span from Myr dynamical times to the 10 Gyr age of the universe. The range in mass starts at the $10^9$ $M_\odot$ scale necessary to resolve galaxies and extends to the $10^{17}$ $M_\odot$ scale comprising a representative sample of the universe. These dynamic ranges, $10^5$ in length, $10^4$ in time, and $10^8$ in mass, are the ideals and no simulation has yet to encompass them all.

Similarly, the physics covers a wide range of processes. In simplest form, simulations model gravity via Newton's law. A collisionless fluid (a.k.a. dark matter) and perhaps a cosmological constant or topological defects are evolved within an expanding universe. To follow the luminous matter, Euler's equations of hydrodynamics governs the evolution of a collisional baryonic fluid (a.k.a. gas). Gas dynamics is more complex and requires the inclusion of artificial viscosity to produce shocks and radiative heating and cooling. Further, the gas will eventually collapse and form stars, converting to a collisionless stellar fluid.

Difficulties arise when physical processes on one scale can affect the state of things on an extremely disparate scales, such as the radiation of a AU scale quasar ionizing gas over Mpc scale volumes. Chief among these effects are star formation, quasar formation, and their associated energy, radiation, and metallicity feedbacks. As the relevant scales are beyond hope of inclusion in the near term, these processes can be included only via analytical or heuristic models.

Considering the scope of the problem, cosmological simulations are one of the largest computing challenges. Cosmological codes have been adapted to nearly every serial, vector, or parallel supercomputer architecture and are an important application for testing new machines and driving further development. Several High Performance Computing and Communication grants have been awarded to groups concentrating on cosmological simulations, and it promises to be a forefront research problem well into the future. At least until the advent of hyper–super–mega–incomparable computers.

## 2. DARK MATTER SIMULATIONS

The basic simulation scheme evolves a collisionless system via gravity. Although other methods are in use, the most common one models the mass as a collection of dark matter particles, also called N-body simulations. Several methods have been developed to solve Poisson's equation for the system; primarily by direct summation of forces between particles (PP, Tree), by calculation of forces on a grid (PM), or by a combination of these techniques ($P^3M$, Tree-PM). The ideas to remember are that grid calculation is quick, but has limited resolution, while the direct summation and combined methods can have arbitrarily high resolution (in theory, though not in practice) at the cost of much, much more computational time.

The simulation method is essentially an initial value problem. One chooses a cosmological model to provide a guess of the conditions in the early universe. A realization of these conditions for a specified volume is then evolved forward in time. Comparing the structures in the simulation to observed structures provides tests of the model. Given a plethora of simulations and observations, of which some of each group are probably in error, no model fits everything. The favored model ends up as the one that is least inconsistent with current observations.

Dark matter simulations are excellent for examining large scale differences between models. Comparisons of simulations to redshift survey results shifted the focus away from the top–down theory of structure formation, as exemplified by the hot dark matter model, and toward the bottom–up, or hierarchical, theory of structure formation, typified by the cold dark matter model. Matter distribution and correlation analysis provide a strong differentiation even in the mildly non-linear regime.

However, in the strongly non-linear regime populated by galaxies, the predictions of dark matter simulations are fuzzy. Figure 1 (all figures are adapted from those in Summers 1993 and Evrard, Summers, & Davis 1994) shows a zoom into a small cluster of galaxies in a dark matter simulation. On the cluster scale, the dark matter halo, as it is called, is quite smooth with little apparent substructure. The one galaxy halo that can be picked out becomes quite bland upon enlargement. A collisionless fluid erases substructure efficiently within collapsed objects. The difficulty of identifying galaxies within clusters as well as the assumptions one must make about how galaxies are distributed relative to the dark matter create large error bars on the small scale predictions of dark matter simulations.

That is not to say that dark matter simulations are no longer useful. If the universe is indeed dominated by collisionless dark matter, then observations that are directly related to the mass are prime candidates for testing models with these simulations. Such observations include the mass function of clusters, cluster density profiles, and gravitational lensing. Because the physics of gravity is computationally simpler and cheaper than hydrodynamics, dark matter simulations are the preferred method of studying these problems.

While some of the numerics of these simulations are not germane to the non-specialist, a few simple checks of resolution can be made by anyone. The largest length scale that can appropriately be addressed in a simulation of box size $L$ is roughly 0.25 $L$. Larger scales, and sometimes even this scale, will be missing power from long wavelengths that are not included. The length resolution scale for grid based simulations is nominally given as one grid cell size (i.e., $L/M$ for $M^3$ grid cells), but the true resolution is 2 – 3 times larger. For the particle direct summation techniques, the nominal resolution scale is the gravitational softening length, the scale over which the gravitational forces are smoothed to avoid point mass interactions. Typically, for $N^3$ particles, the softening is set near 1/16 $L/N$. The true resolution depends on the mathematical form of the force softening, but is invariably 2 – 3 times larger than nominal. The current maximum nominal dynamic ranges are about 500 for grid codes and 4000 for the high resolution particle codes.

Figure 1   The dynamic range in dark matter simulations. Shown here are successive enlargements of a region containing a small cluster of galaxies. The top panels are slices 700 kpc thick and show only one fourth of the particles. Bottom panels are cubical and show all particles.

Linear dynamic range is not the end of the story. The choice of timestep is important because one does not want particles to pass by each other too quickly to interact. The standard choice is to keep the maximum distance a particle will travel down to less than a third of a nominal resolution length. In equation form, it reads

$$\Delta t \leq \frac{1}{3} \frac{\Delta x}{v_{max}}$$

with $\Delta t$ as the timestep, $\Delta x$ the nominal resolution scale, and $v_{max}$ the maximum particle velocity. For a $10^{10}$ year evolution with 10 kpc resolution and simulating 2000 km/s clusters, one needs a timestep $\Delta t \leq 1.2$ Myr or about 8300 timesteps (assuming an extra factor of $\sqrt{2}$ to cover the tail of the velocity distribution). The dynamic range in mass has a nominal resolution for particle simulations of simply one particle mass while true resolution is of order ten particles. In grid simulations which use particles, note that the nominal mass resolution is the mass within an entire grid cell, and cannot be subdivided into particles within a cell. Any simulations which deviate strongly from the above ideas should justify their choices.

## 3. EULERIAN HYDRODYNAMICS

To include a baryonic component, one may model it as gaseous fluid on a fixed (Eulerian) grid. These codes (PPM, TVD), also called finite difference methods or computational fluid dynamics, solve the equations of hydrodynamics on the grid and couple the system to a gravity solver. The standard choice of gravity solver is an N-body code that also computes on a grid (PM), thereby providing similar resolution scales for both processes.

The best applications of Eulerian codes are examinations of the gaseous component of the universe. Such applications include studies of the X-ray gas in clusters, Lyman $\alpha$ clouds, the inter-galactic medium, and the relative distributions of gas and dark matter. Work on X-ray clusters has determined that the COBE normalized cold dark matter model produces too many and too luminous clusters, while both the mixed dark matter model (some cold, some hot) and a cold dark matter model with a cosmological constant can fit the observations adequately (e.g., Bryan et al. 1994; Kang et al. 1994). Other research has suggested that Lyman $\alpha$ clouds are not the individual collapsed objects their moniker would imply, but instead may be gas in an earlier stage of collapse, that within caustics (e.g., Cen et al. 1994). These simulations should continue to provide new ideas on gas from voids to moderate overdensities.

And the future holds considerable promise. Computational fluid dynamics is a large and well sponsored field, having military, aerospace, and automotive applications. Cosmology has tapped only a portion of these resources. Most promising are the techniques which adapt the grid to the distribution of the fluid by adding more cells in high density regions. These techniques increase the spatial dynamic range as the simulation progresses and will allow one to follow the collapse process considerably further.

Dynamic range considerations for hydrodynamic simulations have only a few differences compared to those discussed previously. Grid based methods have the same nominal / true length resolution of 1 / 2–3 grid cells with mass resolution determined by the grid scale. Particle based methods (discussed in §4 below) have their resolution set by the requirement that the smoothing procedure should encompass a scale of about 30 particles. Thus, true mass resolution is around 30 particle masses and the spatial resolution scale is variable – larger in low density regions and smaller in high density regions. Nominal length resolution is given by the minimum smoothing length (again, true is 2–3 times larger), which is often fixed to be near the gravitational softening scale. The current maximum nominal dynamic ranges are about 500 for grid codes (limited by RAM size) and 2000 for particle codes (limited by CPU cycles).

The other major watchpoint for hydro codes is the physics which is not included in the model. "What is missing?" is an important question to ask of the results. Will ionizing radiation from quasars slow or prevent galaxy collapse (Babul & White 1991)? Does line emission from metals contribute to the X-ray flux in the relevant bandpass (Cen et al. 1995)? Is the energy injected from supernovae a significant component in heating an intra-cluster medium (Lowenstein & Mushotzky 1995)? These and other questions will continue to shape and refine interpretation of simulation results. The bottom line is that, due to physics not included, some measures are meant to be considered as underestimates or as upper limits.

## 4. LAGRANGIAN HYDRODYNAMICS

The process of increasing resolution in collapsing areas happens naturally when one switches to a LaGrangian formulation (i.e., the cells move with the fluid). Three dimensional LaGrangian grid systems can produce about a factor of ten increase in range, but then break down when the grid becomes strongly distorted. The more successful approach in cosmology has been the technique of smoothed particle hy-drodynamics (SPH), in which gas particles act both as tracers of the fluid and as calculation points for the hydrodynamics. SPH uses kernel smoothing with variable smoothing lengths to cover a wide range of scales. As a particle based method, SPH joins cleanly with the high resolution particle algorithms for gravity ($P^3$MSPH, TreeSPH).

Figure 2   The dynamic range in SPH simulations. The same simulation and the same panels are used as in Figure 1, but now the baryonic gas particles are displayed.

A direct comparison to the dark matter models is provided in Figure 2. On the larger scales, dark matter and gas distributions are similar because gravity is the dominant force and the collapse is essentially pressureless until a shock is reached. Within the collapsed regions,

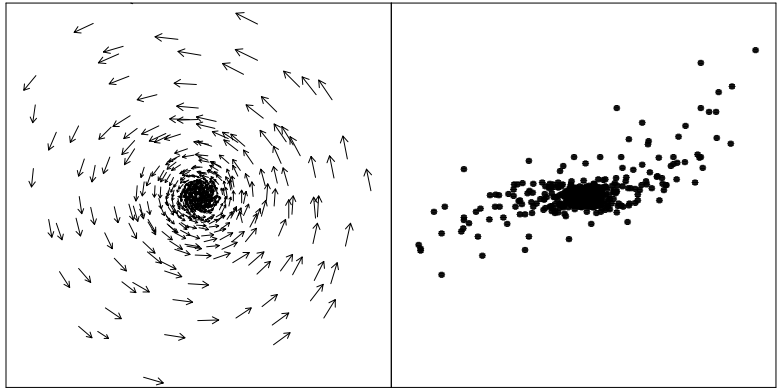

Face On          Edge On

Figure 3   An SPH disk galaxy. All baryon particles with temperature less than $3 \times 10^4$ K are plotted within a 60 kpc cube.

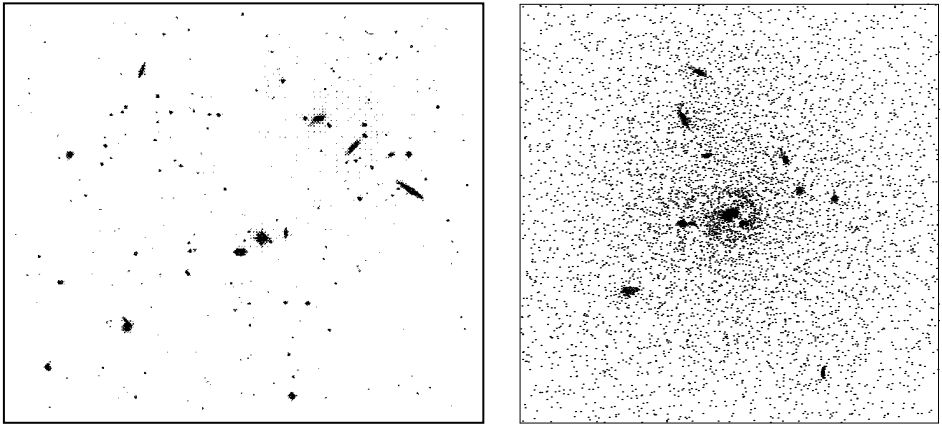

Group 11 55 +2526          Simulated Group

Figure 4   The left panel shows a scanned finding chart from the Palomar survey for the group at 11 55, +25 26. The right panel shows a group from an SPH simulation. Linear scales are similar, roughly 500 kpc across.

however, the baryons have cooled and condensed to galaxy scales, and these simulated galaxies populate the cluster. Many of these objects show rotationally supported disk structures (Figure 3). Groups of these simulated galaxies can show a striking resemblance to observed galaxy groups (Figure 4).

SPH simulations are well suited to studies of highly collapsed ob-jects. Galaxy formation studies can delve into the formation processes, formation epochs, merging histories, (rough) morphologies, and other characteristics as a function of redshift (e.g., Katz, Hernquist, & Weinberg 1993; Evrard et al. 1994). One can also follow the aggregation of larger structures, gaining insight into the formation processes, structure, dynamics, and gas content of groups and clusters (e.g., Pildis, Bregman, & Evrard 1995; Navarro, Frenk, & White 1995; Frenk et al. 1995). When sufficient dynamic range is available, one can directly compare the statistics of simulated galaxies against large scale structure measures from redshift surveys. As a highlight of my recent work, the next section describes a premier example of SPH simulations.

## 5. AN IMAX FILM PREVIEW

In May 1996, the film *Cosmic Voyage* will debut at the five story tall IMAX theatre in the National Air and Space Museum. This film will take the viewer on a journey through the grand sweep of space and time: from the realms of sub-atomic particles to the largest scales of the known universe, from the present day back to the earliest seconds after the Big Bang. Members of the Grand Challenge Cosmology Consortium are providing the scientific basis for a four minute sequence in *Cosmic Voyage* that depicts the formation of structure in the universe. Beginning shortly after the Big Bang, the sequence follows the expansion of the universe, the gravitational collapse of structure, the formation of galaxies, and a collision of of two grand design spiral galaxies. This sequence is the only part of the film that is direct scientific visualization: based on data from scientific simulations rather than artist's conceptions.

The three parts of the sequence are being computed separately. Greg Bryan (Illinois) and FJS are working with the computer graphics experts at PIXAR to create a visualization of the early universe based on standard power spectrum formulae. The challenge is to coherently visualize and animate the tiny fluctuations in the density field within a three dimensional, expanding, and evolving universe. In the second part, FJS is creating a simulation of galaxy formation with unprecedented dynamic range. A dramatic zoom shot will cover scales from the filaments that connect clusters of galaxies down to the collapse of an individual galaxy. The third section features Chris Mihos' (Santa Cruz) simulation of two galaxies colliding and merging with intricate tidal streams of stars and gas, as well as violent bursts of star formation. Depth scale will be added by continuing a portion of FJS' simulation as a background to the merger.

Computing for IMAX resolution pushed the simulation art to the limits. The galaxy formation simulation (FJS) is a four million particle SPH run that required several CPU weeks on a dedicated 8 processor SGI Power Challenge. It produced 110 GB of data. The merger simulation also used SPH techniques to follow a quarter million particles. 750 CPU hours of computation on a Cray C90 produced over 65 GB of data. Both simulations are the largest of their kind performed to date. Visualizations are being created by a team at NCSA using proprietary software written at PIXAR. Today, I can show a short test video of the galaxy formation sequence which conveys a sense of the enormous amount of structure in the simulations and the stunning images that will be in the IMAX film.

[Insert mondo-cool video clip here.]

## ACKNOWLEDGEMENTS

Support for this work was provided by NSF grant AST-8915633 and NASA grant NAGW-2367. Computing resources provided by the San Diego Supercomputing Center, the National Center for Supercomputing Applications, and the Pittsburgh Supercomputing Center are gratefully acknowledged. Special thanks to V. Trimble for last minute adjustments and VCR procurement.